\documentclass{kluwer}    

\usepackage[]{graphicx}

\begin{document}                                                                                   
\begin{article}
\begin{opening}        
\title{The properties of Low Surface Brightness galaxies}
\author{D.\surname{ Monnier Ragaigne \email{delphine.ragaigne@obspm.fr}}}
\author{W.\surname{ van Driel}}
\author{C.\surname{ Balkowski}}
\institute{Observatoire de Paris, GEPI,  Meudon, France }
\author{S.\surname{Boissier}}
\institute{Institute of Astronomy, University of Cambridge, UK}
\author{N.\surname{Prantzos}}
\institute{Institut d'Astrophysique de Paris, Paris, France}

\runningauthor{ D. Monnier Ragaigne et al.}
\runningtitle{The properties of LSB galaxies}

\begin{abstract}
A description is given of the samples of Low Surface Brightness galaxies
(LSBs) used for comparison with models of their chemical and spectro-photometric
evolution (Boissier et al., this Volume). These samples show the large variation
and scatter in observed global properties of LSBs, some of which cannot be
modeled without adding starbursts or truncations to their star formation history.
\end{abstract}
\end{opening}

\section{Introduction}
In the past few decades, the existence has been shown of galaxies with a blue central 
disc surface brightness well below the Freeman value of 
$\mu_{B,0}$=21.65 mag arcsec$^{-2})$, which is typical of the average previously catalogued
``classical'' High Surface Brightness  spirals (HSBs). 
Particularily in the last decade a considerable body of observational data
has been collected on this class of Low Surface Brightness galaxies (LSBs),
which may turn out to be crucial to studies of galaxy formation and evolution and of
the `cosmic' chemical evolution,  if (as suggested by O'Neil \& Bothun 2000) they represent the 
majority of galaxies. 

Although there is no unambiguous definition of an LSB galaxy, we adopted as the limit 
between HSBs and LSBs a blue central disc surface brightness value of
$\mu_{B,0}$=22 mag arcsec$^{-2}$, a commonly used criterion.

Though LSBs show a large variety of observed properties (see below), their obvious 
interest has motivated various theoretical studies towards an understanding
of their nature. In a recent series of papers, the chemical and spectro-photometric
evolution of ``classical'' HSB spiral galaxies was computed (for various galactic masses
and angular momentum) and compared quite successfully to various types
of observational quantities like colours, spectrum, star formation efficiency and gas fraction 
as well as to their Tully-Fisher relationship (Boissier \& Prantzos 1999,2000; Boissier et al. 2001;
see also Prantzos et al.,  this Volume).
These models have now been extended to cover LSBs (Boissier et al. 2002, and this Volume),  
based on the suggestion of Dalcanton et al. (1997) that LSBs have a 
larger angular momentum than HSBs, as used also by Jimenez et al. (1998) in their models.

\section{The diversity of LSB galaxies}
Observations have shown a large variety of LSB galaxies, from dwarfs to giants
and from ``blue'' to the red edge of galactic colours, which indicates that they form a 
heterogeneous family.
In our models we explore a possible link between HSB and LSB discs, without an
{\it a priori} idea of which of the LSBs should actually be linked to HSBs. 
We therefore have to consider the whole range of LSBs, which
can be achieved by the use of the following data sets available in the literature. 
For each sample at least detailed surface photometry, colours and 21 cm H\,{\sc i} line 
spectra should be available.

{\it O'Neil et al. LSBs sample}:\, 
The data from O'Neil et al. (1997a,b) provide
information on the surface brightness, scale-length, absolute magnitude of the disc
and colours for 127 LSBs,
80 \% of which are well fitted by an exponential profile, justifying the use of a 
disc-model for studying LSBs. These data showed that a large fraction of LSBs 
were redder than previously thought. H\,{\sc i} data for subsets 
are given in O'Neil et al. (2000a) and Chung et al. (2002).
Some objects are members of clusters (Pegasus and Cancer). 

{\it De Blok et al. LSBs sample}:\, 
For 21 late-type field LSBs de Blok et al. (1995) published disc scale-lengths, absolute magnitudes 
and colours, while for subsets interferometric H\,{\sc i} data are given in 
de Blok et al. (1996), abundances in de Blok \& van der Hulst (1998a), CO line upper limits
in de Blok \& van der Hulst (1998b) and H$\alpha$ rotation curves in 
McGaugh et al. (2001).
Their gas content is also available from van den Hoek et al. (2000).

{\it Bulge-dominated LSBs sample}:\, 
Although most LSBs seem to be late-type objects without any
apparent bulge, some of them clearly do have bulges, and we should not 
exclude these from our analysis, to avoid introducing a bias against the most
evolved LSB discs. For this reason, we also use the sample of 20 LSBs from 
Beijersbergen et al. (1999). 

\begin{figure}
\includegraphics[width=9.5cm,angle=-90]{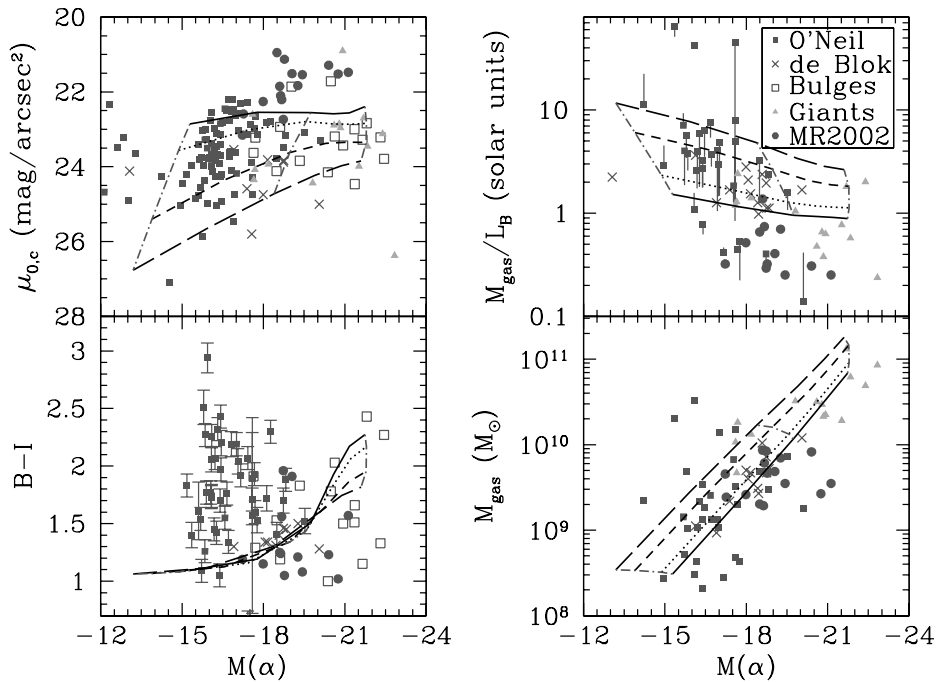}
\caption{\label{figone}
Some plots showing the large diversity and scatter in the properties
of LSB galaxies; for the superimposed
model grids, we refer to Boissier et al. (this Volume). Shown as function of the 
absolute blue magnitude of the disc, M($\alpha$), are: 
the blue central disc surface brightness (top left), total $B-I$ colour (top right), 
total H\,{\sc i} mass (bottom left) and H\,{\sc i} mass-to-blue luminosity ratio 
(bottom right). Note that the infrared-selected sample (MR2002) is on average 
less LSB in the $B$-band than the others and contains objects which are in fact 
HSBs,  according to our optical selection criterion of $\mu_{B,0}$=22 mag arcsec$^{-2}$.}
\end{figure}

{\it Giant LSBs sample}:\, 
We used the  data of Matthews et al. (2000), which provides surface photometry 
and H\,{\sc i} spectra for 16 giant LSB galaxies, with high luminosities (10$^{10}$ L$_{\odot}$)
and large scale-lengths ($>6$ kpc). Though LSB Giants are relatively rare, they
may play an important role in our investigation of the link between HSBs and
LSBs because of their similar sizes/velocities (note that there are many more HSBs with 
large line widths than with small ones, like for LSBs).

{\it Infrared-selected LSBs sample}:\, 
A new sample of LSBs selected from  the 2MASS near-infrared survey
(Monnier Ragaigne et al. 2002) is used as well. 
Though these 4,000 objects were selected on their low $K_s$ band central
disc surface brightness, $\mu_{K,0}$$<$18 mag arcsec$^{-2}$, they turned out to
be bluer than expected and their $B$ band central disc surface brightness lies
in the range of $\sim$21-22.5  mag arcsec$^{-2}$, around the optical selection
criterion for LSBs we adopted.
Here, we use the subset of 20 galaxies for which we presently have available
$BVRI$ surface photometry, H\,{\sc i} spectra and $JHK$ photometry.
This sample allows the exploration of the link between optically HSB and
LSB galaxies, as can be seen in Figure 1, where has been labelled as MR2002.

Because the data come from various sources, we have homogenised them
for a proper comparison with our disc models, as follows:
(1) all distance-dependent quantities were corrected to 
a Hubble constant of 65 km/s/Mpc;
(2)  the data were corrected to face-on, i=0$^{\circ}$, assuming LSB discs
to be optically thin, which seems justified by the fact that their colours are
independent of inclination (e.g. O'Neil et al. 1997a);
(3) the absolute total disc magnitudes, $M(\alpha)$, are deduced from an
exponential fit to the exponential profiles (e.g. O'Neil et al. 1997a), and
(4) the total gas mass is obtained by correcting the H\,{\sc i} mass
for the Helium fraction ($M_G$/$M_{HI}$=1.4). The molecular component
in LSBs is generally negligible, though a number have now been detected
in CO (Matthews \& Gao 2001; O'Neil et al. 2000b, and references therein).

The large diversity between, and among, the LSBs in these samples is obvious 
from the plots shown in Figure 1. Though most properties can be fitted with our 
''simple'' models, which have a smoothly changing star formation rate over time,   
in order to explain the wide range in colours, especially the red objects in the 
O'Neil et al. sample, star bursts and truncations need to be added to the star formation
history of the ``simple'' models (Boissier et al., this Volume).

\end{article}


\vspace{-0.5cm}

\begin{thebibliography}{}
\bibitem{} Beijersbergen M., de Blok W. J. G.,  van der Hulst J. M., 1999, A\&A, 351, 903 
\bibitem{} Chung A., van Gorkom J. H., O'Neil K., Bothun G. D., 2002, AJ,  123, 2387 
\bibitem{} Dalcanton J.,  Spergel D., Summers F., 1997, ApJ,  482, 659 
\bibitem{} de Blok W. J. G., van der Hulst J. M., Bothun G. D., 1995, MNRAS, 274, 235 
\bibitem{} de Blok W. J. G., McGaugh S. S., van der Hulst J. M.,  1996, MNRAS, 283, 18
\bibitem{} de Blok W. J. G., van der Hulst J. M.,  1998a, A\&A, 335, 421
\bibitem{} de Blok W. J. G., van der Hulst J. M.,  1998b, A\&A, 336, 49
\bibitem{} Boissier S., Prantzos N., 1999, MNRAS, 307, 857 (BP99)   
\bibitem{} Boissier S., Prantzos N., 2000, MNRAS, 312, 398 (BP2000) 
\bibitem{} Boissier S., Boselli A., Prantzos N., Gavazzi G., 2001, MNRAS, 321, 733
\bibitem{} Boissier S., Monnier Ragaigne D.R., Prantzos N., van Driel W., Balkowski C.
  2002, MNRAS, submitted
\bibitem{} Jimenez R., Padoan P., Matteucci F., Heavens A., 1998, MNRAS, 299, 123 
\bibitem{} Matthews L. D., van Driel W., Gallagher J. S., 1998, AJ, 116, 2196
\bibitem{} Matthews L. D., Gao Y., 2001, ApJ, 549, L191
\bibitem{} Monnier Ragaine D. R., van Driel, W., Schneider S.E., Jarrett T.H., Balkowski, C., 2002,
 A\&A, submitted
\bibitem{} McGaugh S. S., Rubin V., De Blok W. J. G., 2001, AJ, 122, 2381
\bibitem{} O'Neil K., Bothun G. D., Cornell M., 1997a, AJ, 113, 1212 
\bibitem{} O'Neil K., Bothun G. D., Schombert J., Cornell M., Impey, C., 1997b, AJ, 114, 2448 
\bibitem{} O'Neil K., Bothun G. D. 2000, ApJ, 529, 811
\bibitem{} O'Neil K., Bothun G. D., Schombert J., 2000a, AJ, 119, 136 
\bibitem{} O'Neil K., Hofner P., Schinnerer E., 2000b, ApJ, 545, L99
\bibitem{} van den Hoek L., de Blok W. J. G., van der Hulst J. M., de Jong T., 2000, A\&A, 357, 397 

\end{thebibliography}
\end{document}